\documentclass[floatfix,amsmath,amssymb,twocolumn,aps,pra,superscriptaddress]{revtex4-2}

\usepackage{graphicx}% Include figure files
\usepackage{dcolumn}% Align table columns on decimal point
\usepackage{bm}% bold math
\usepackage{hyperref}% add hypertext capabilities
\usepackage[normalem]{ulem} % strikethrough text
\usepackage{color}
\usepackage[dvipsnames]{xcolor}
\usepackage{natbib}

\begin{document}

\title{Quantum reservoir computing in atomic lattices}

\author{Guillem Llodrà}
 \email{guillemllodra@ifisc.uib-csic.es}
\affiliation{%
 Institute for Cross-Disciplinary Physics and Complex Systems (IFISC) UIB-CSIC, Campus Universitat Illes Balears, Palma de Mallorca, Spain.
}
\author{Pere Mujal}
\affiliation{
ICFO-Institut de Ciencies Fotoniques, The Barcelona Institute of Science and Technology, 08860 Castelldefels (Barcelona), Spain
}
\author{Roberta Zambrini}
 %\email{roberta@ifisc.uib-csic.es}
\affiliation{%
 Institute for Cross-Disciplinary Physics and Complex Systems (IFISC) UIB-CSIC, Campus Universitat Illes Balears, Palma de Mallorca, Spain.
}
\author{Gian Luca Giorgi}
\email{gianluca@ifisc.uib-csic.es}
\affiliation{%
 Institute for Cross-Disciplinary Physics and Complex Systems (IFISC) UIB-CSIC, Campus Universitat Illes Balears, Palma de Mallorca, Spain.
}

\date{\today}

\begin{abstract}
Quantum reservoir computing (QRC) exploits the dynamical properties of quantum systems to perform machine learning tasks. We demonstrate that optimal performance in QRC can be achieved without relying on disordered systems. Systems with all-to-all topologies and random couplings are generally considered to minimize redundancies and enhance performance. In contrast, our work investigates the one-dimensional Bose-Hubbard model with homogeneous couplings, where a chaotic phase arises from the interplay between coupling and interaction terms. Interestingly, we find that performance in different tasks can be enhanced either in the chaotic regime or in the weak interaction limit. Our findings challenge conventional design principles and indicate the potential for simpler and more efficient QRC implementations tailored to specific tasks in Bose-Hubbard lattices.
\end{abstract}

\maketitle
\normalsize
\section{Introduction}
Despite significant progress in quantum computing, the quest for large-scale, fault-tolerant, and universal implementations suited for practical problems remains an ongoing challenge \cite{arute2019quantum,zhong2020quantum,madsen2022quantum}. 
The current noisy intermediate-scale quantum (NISQ) devices \cite{preskill2018quantum,bharti2022noisy}, characterized by a small number of quantum units and moderately high error rates, are not yet capable of achieving the practical purposes of the pioneering and promising algorithms 
\cite{Shor,Grover} that boosted the development of current conventional gate-based technologies in qubit systems. NISQ devices fostered advances in machine learning (ML), mainly in variational quantum algorithms (VQA) \cite{CerezoReview2021}. Nevertheless, the training process of these parameterized circuits presents significant challenges that limit their successful implementation, as evidenced by the phenomenon of barren plateaus \cite{LaroccaBarrenPlateaus,SanniaBarrenPlateaus}.
 
Together with the advancement of quantum computing, quantum simulators have emerged, inspired by Feynman's concept of emulating the behavior of a quantum system of interest by utilizing an equivalent one \cite{Feynman1982}. 
The notable progress obtained in the control and manipulation of quantum systems, coupled with a significant reduction in hardware requirements, has resulted in the %advent 
development of quantum simulators. Currently, these are predominantly implemented in analog systems \cite{PRXQuantum.2.017003,daley2022practical,trivedi2024quantum}, although proposals for digital gate-based ones also exist  \cite{Fauseweh2024}.
In particular, one successful approach to quantum simulation is the use of Bose-Einstein condensates (BECs) \cite{yamamoto2016bose,RevModPhys.80.885}. By exploiting the unique properties of BECs, a wide range of quantum phenomena have been simulated, from superfluidity \cite{bogoliubov1947theory, landau2018theory, tao2023observation, chauveau2023superfluid} and superconductivity \cite{heyl2022vortex, randeria2014crossover}, to quantum phase transitions and many-body localization \cite{barbiero2015out, see2022many}, and topological gauge theories \cite{Frolian2022}.
Additionally, cold atom experiments with quantum gas microscopes have reached an unprecedented control over the preparation, manipulation and read-out of highly isolated systems of large ensembles of both bosonic and fermionic atomic species \cite{Buob_Tarruell2024,Gross_Bakr2021,Eliasson_Sherson2020,Kuhr_2016,Preiss_Greiner2015,Bakr_Greiner2009}. 

A less explored avenue for these quantum simulators is in the context of unconventional and neuromorphic computing \cite{adamatzky2007unconventional,markovic2020quantum,mujal2021opportunities},  
where these platforms can be adapted and exploited to implement novel algorithms and protocols. One promising application, inspired by recurrent neural network, at the intersection of unconventional computing, machine learning, and analog systems, is quantum reservoir computing (QRC) \cite{fujii2017harnessing, mujal2021opportunities}. Reservoir computing is a supervised machine-learning technique allowing for in-memory processing and easy training, where the reservoir is kept fixed and only the output layer is trained, usually using a simple linear regression, that allows for efficient training and multitasking  \cite{tanaka2019recent,Nakajima_2020_physical,nakajima_book}. 
Beyond inheriting the advantages of classical settings, QRC approach targets either classical or quantum temporal information processing, enhanced by the large state dimension and capabilities of quantum physical reservoirs, even in NISQ devices \cite{mujal2021opportunities}.

The performance of QRC  depends on the echo state and fading memory \cite{jaeger2001echo,grigoryeva2018echo,sannia2024dissipation,nokkala2021gaussian,RodrigoJuanPablo2023} properties, as well as a suitable nonlinear input-output transformation \cite{mujal2021analytical,govia2022nonlinear}, 
associated to the proper hyperparameters choice \cite{martinez2021dynamical}.
The potential of QRC  has been demonstrated through their information processing capacity \cite{martinez2023information} and other memory tasks \cite{fujii2017harnessing,nokkala2024retrieving}, and in forecasting, as dynamical system \cite{fry2023optimizing} chaotic  \cite{tran2020higher} or financial \cite{xia2023configured, domingo2023anticipating,kutvonen2020optimizing} time series predictions. 
Examples of static tasks, that can be realized also in absence of internal memory --a setting known as Quantum Extreme Learning Machine (QELM)-- are entanglement detection \cite{ghosh2019quantum} or ground-state finding \cite{mujal2022quantum} among others \cite{mujal2021opportunities}. Experimental realizations of QRC and QELM have already been reported in nuclear magnetic resonance \cite{negoro2018machine}, superconducting qubits \cite{Chen2020,Suzuki2022,Kubota2023,Yasuda2023,senanian2024microwave}, photonic setups \cite{Paternostro2024}, and Rydberg atoms \cite{Yelin2022, Yelin2024}.

In this work,  we consider a one-dimensional Bose-Hubbard model as a quantum reservoir, focusing on two key questions: optimizing the operational regime and understanding the role of disorder, often considered an essential ingredient for QRC. To address these, we numerically investigate the dynamics of the Bose-Hubbard model for different QRC benchmarking tasks 
in the crossover between superfluid and Mott-insulator phase \cite{Fisher1989}, experimentally tested in \cite{Greiner2002,Bakr2010,Sherson2010,Tomita2017}.
Recently, it was shown that signatures of quantum chaos arise in a transition regime in the Bose-Hubbard model \cite{pausch2021chaos, pausch2021chaos_NJP}. Following the results reported in \cite{martinez2023information}, we aim to address the relative performance of QRC in this quantum chaos regime with respect to the limits where interactions or hopping dominate.

We also address the role played by disorder and symmetries in QRC substrates. 
While simple periodic structures like one-dimensional chains exhibit limited performance, breaking translational symmetry in (classical and quantum) reservoirs is known to significantly enhance processing performance  \cite{dale2021reservoir, mallinson2023reservoir, griffith2019forecasting}.  Actually, with the exception of  (problem related) symmetries that can prove beneficial \cite{barbosa2021symmetry},   generic tasks lack known symmetries and these are detrimental \cite{herteux2020breaking, flynn2021symmetry,martinez2021dynamical}. In our study, we show that QRC can be implemented using more straightforward configurations not requiring to tune the Bose-Hubbard lattice towards inhomogeneous configurations, which may facilitate the conduct of practical experiments with actual quantum devices.

The paper is organized as follows. We first introduce the Bose-Hubbard model and discuss its phases and quantum chaos in section \ref{sec.BHmodel}. Section \ref{secQRC} contains all the details about the quantum reservoir computing algorithm. Then, in section \ref{sec:Reservoir performance} there is an analysis on the performance depending on the dynamical regime for different benchmarking tasks. Section \ref{sec:measurement_effects} addresses the impact of finite measurements on the previous tasks. In section \ref{sec:topologies}, we focus on the performance across different topologies, highlighting their influence on learning capabilities. Finally, the conclusions are presented in section \ref{sec.conclusions}.

\section{Bose-Hubbard Model}
\label{sec.BHmodel}
We consider a one-dimensional chain of interacting bosons as a quantum reservoir computer, described by the  Bose-Hubbard Hamiltonian \cite{Fisher1989,kuhner2000one,ejima2011dynamic,RevModPhys.80.885}:
\begin{equation}
H=-J\sum_{j=1}^{N}(b^\dagger_j b_{j+1} + \mathrm{h.c.}) + \frac{U}{2}\sum_{j=1}^{N}n_j(n_j-1),
\label{eq:bose-hubbard hamiltonian}
\end{equation}
where $b^\dagger_j (b_j)$ are the bosonic creation (annihilation) operators at site $j$, $n_j$ is the number operator, $J$ is the tunneling parameter characterizing the boson mobility between nearest-neighbor sites, and $U$ is the on-site repulsive ($U > 0$) interaction strength. 
The Bose-Hubbard model has three important symmetries: $\mathbb{Z}_2$, translational --under periodic boundary conditions-- and conservation of the total boson occupation number. While these symmetries simplify the model, we anticipate that they can be detrimental to fully exploiting the capabilities of quantum reservoir computing limiting its expressivity (see Section \ref{sec:topologies}).  
In this work, we consider mainly a chain with open boundaries (where $b_{N+1}=0$), but also periodic boundaries (where $b_{N+1}=b_{1}$),  the disordered chain with inhomogeneous (and random) tunneling couplings ($J\rightarrow J_{j,j+1}$), and all-to-all network.

The interplay between the Hamiltonian terms leads to a rich variety of regimes \cite{Bakr2010, Stoferle2004, Spielman2007,RevModPhys.80.885, Kohl2005,Greiner2002,Sherson2010}. In the limit of strong on-site interaction $(U\gg J)$, the ground state of the system is in the Mott-Insulator (MI) phase \cite{RevModPhys.80.885,Greiner2002}, where each site is occupied by a fixed number of bosons, and the tunneling between sites is suppressed. In the opposite limit, $(U\ll J)$ the ground state of the system is in the superfluid (SF) phase \cite{RevModPhys.80.885,Greiner2002}, where the bosons can move freely through the chain. 
Recently, looking at the full energy spectrum of the Hamiltonian in the region where $U \sim J$, signatures of a dynamical phase transition to chaos were reported \cite{pausch2021chaos, kolovsky2004quantum}. It was shown that the distribution of energy level spacings follows a Wigner-Dyson distribution \cite{oganesyan2007localization}.  In the following, we will mainly be concerned with the dynamical phases of the system, which are candidates to play a determinant role in the reservoir capabilities \cite{martinez2021dynamical}.
With abuse of language by mixing nomenclatures from quantum phase transitions and dynamical phase transitions,  and for the sake of simplicity, we will refer to the different parameter regions as the MI phase $(U\gg J)$, the chaotic phase $(U\sim J)$ and the SF phase $(U\ll J)$.

% Paragraph about Quantum Chaos, RMT and GOE ensembles
Among tools to identify the presence of quantum chaos, random matrix theory (RMT) (originated from understanding energy level statistics in complex nuclei \cite{RevModPhys_RMT_nuclei}) has become a fundamental framework in the analysis of quantum many-body spectra \cite{d2016quantum}. Let us consider the distribution of level spacing, $p(s)$, between consecutive energy levels, $s_n=E_{n+1}-E_n$. According to RMT,  when $p(s)$ follows a Poisson distribution, the system tends to be integrable, while a Wigner-Dyson distribution signifies chaotic behavior. Different underlying symmetries of the system correspond to different ensembles of random matrices, with the Gaussian orthogonal ensemble (GOE) being particularly pertinent for studying systems like the Bose-Hubbard model \cite{pausch2021chaos, pausch2021chaos_NJP}. Furthermore, methods like the gap ratio, $r_n=min(s_n/s_{n-1}, s_{n-1}/s_n)$, introduced by Oganesyan and Huse \cite{oganesyan2007localization} simplify the calculations and gave a clear indicator of ergodicity, $\langle r \rangle_{GOE}\approx 0.5359$, and integrability, $\langle r \rangle_{\textnormal{Poisson}} \approx 0.3863$ \cite{atas2013distribution, giraud2022probing}. This indicator was used in Refs. \cite{martinez2021dynamical,palacios2024role} as a marker of the QRC regimes and related capabilities for the transverse Ising model.
 
The transition to chaos in the Bose-Hubbard model has recently been analyzed by Pausch  \textit{et al.} \cite{pausch2021chaos, pausch2021chaos_NJP}. In these works, the authors examined the eigenvalue spacing, as characterized by $\langle r\rangle$, and the eigenstate structure, through the so-called generalized fractal dimension (GFD). These measures allow for the distinction between localization, multifractality (extended nonergodic), and ergodicity. Given a state $|\psi\rangle=\sum_\alpha \psi_\alpha |\alpha\rangle $ (where $\{ |\alpha\rangle\}$ is a complete basis that spans a Hilbert space of dimension $\mathcal{N}$), among all the finite-size GFDs, we will consider the quantity $\tilde{D}_1$, known as the \textit{information dimension} \cite{lindinger2019many}, which determines the scaling behavior of the Shannon information entropy:  $\tilde{D}_1 = -(\ln{\mathcal{N}})^{-1}\sum_\alpha |\psi_\alpha|^2\ln{|\psi_\alpha|^2}$.
The thermodynamic limit of this quantity, $D_1=\lim_{\mathcal{N}\to\infty}\tilde{D}_1$  provides insight into the eigenstate distribution in the Hilbert space. For $D_1=1$, the state is spread uniformly over the entire Hilbert space, which is a signature of ergodicity. On the other hand, $D_1=0$ corresponds to a perfectly localized state. For intermediate values, $0 < {D}_1 < 1$ the system exhibits multifractal behavior, characterized by a complex combination of localization and ergodicity. In such cases, the wavefunction occupies a fractal subset of the Hilbert space, extending non-ergodically as discussed in \cite{lindinger2019many}.

\begin{figure}[t] % Place the figure at the top of the page
    \centering
    \includegraphics[width=0.48\textwidth]{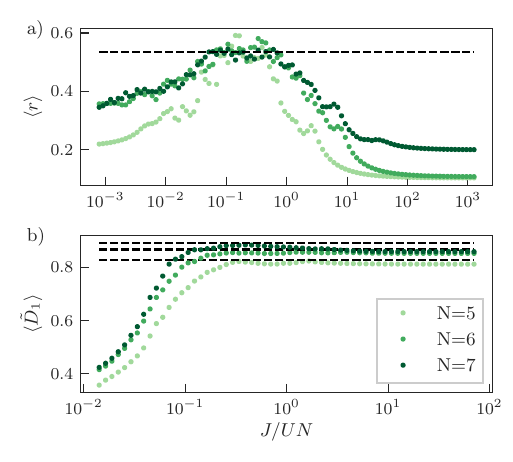} % Adjust the width as needed
    \caption{(a) Distribution of the average level spacing ratios $\langle r \rangle$ over the inner 70\% of the eigenenergies as a function of $J/UN$ for one-dimensional open with parity ($\pi = -1$) at unit-filling ($N=N_e$) being $N$  ($N_e$) the number of sites (excitations) in the chain. (b) Evolution of $\langle \tilde{D}_1 \rangle$ as a function of $J/UN$. The data is averaged from the amplitudes $\psi_{\alpha}$ of the 100 eigenvectors closest to rescaled energy, $\epsilon = (E-E_{min})/(E_{max}-E_{min}) = 0.5$ where $\alpha$ represents the Fock basis. The black line corresponds to the expected results for $\langle r \rangle_{GOE}\approx0.5359$ and $\langle \tilde{D}_1 \rangle = (H_{\mathcal{N}/2}-2+\ln 4)(\ln \mathcal{N})^{-1}$ \cite{pausch2021chaos} being $\mathcal{N}=60, 226, 848$ for $N=5, 6, 7$, respectively.}
    \label{fig:ratio_and_gfd}
\end{figure}

The behavior of both $\langle r\rangle$ and $\langle \tilde{D}_1 \rangle$ (where the basis $\{ |\alpha\rangle\}$  corresponds to the Fock basis in this context) is reproduced in Fig. \ref{fig:ratio_and_gfd} for Bose-Hubbard reservoirs with chain sizes from $N=5$ to $N=7$. One can see that multifractality is a sensitive tool for characterizing the emergence of chaos, with plateau values closely matching the GOE values, as indicated by the black lines in Fig. \ref{fig:ratio_and_gfd}. We will see that this behavior has important consequences once we look at the QRC capabilities of the model in the different dynamical regimes.

\section{Quantum Reservoir Computing based on the Bose-Hubbard model}
\label{secQRC}
\begin{figure}[t] % Place the figure at the top of the page
    \centering
    \includegraphics[width=0.48\textwidth]{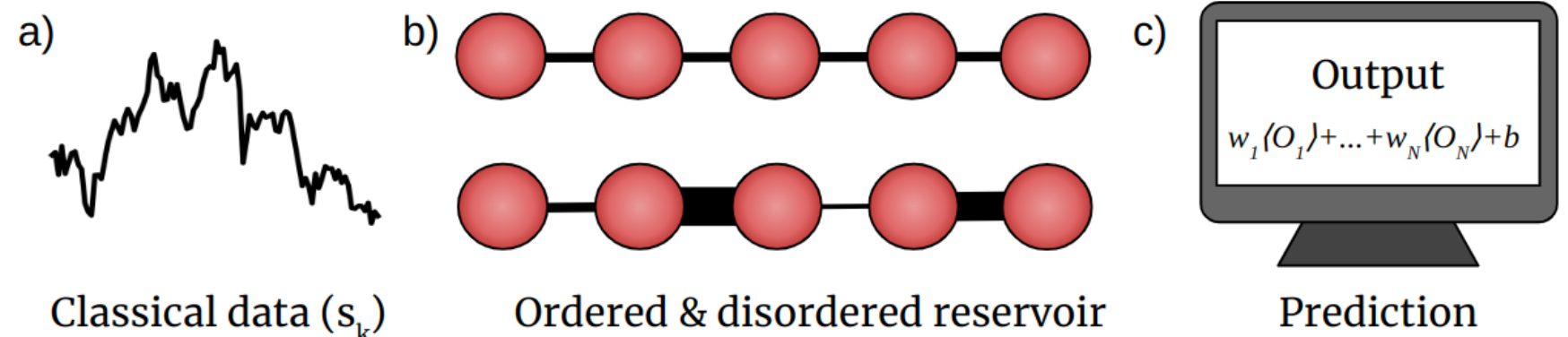} % Adjust the width as needed
    \caption{Illustration of the Quantum Reservoir Computing algorithm steps. a) Classical data ($s_k=[0, 1]$) is fed into the first site of the system. b) The classical data is processed through a one-dimensional quantum reservoir. In this work, we will compare the performance of a homogeneous and heterogeneous system. c) Output predictions are generated based on the expected values of the quantum system.}
    \label{fig:illustration_QRC}
\end{figure}

Let's briefly introduce the machine learning approach of QRC that will be realized with a Bose-Hubbard chain substrate. This is composed of three fundamental layers, as illustrated in Fig. \ref{fig:illustration_QRC}. A (classical or quantum) input signal is fed into the reservoir. In this case, a one-dimensional chain of atoms processes the received information. Then, meaningful features from the reservoir are extracted at the output layer as a set of observables.
Suppose our input consists of a sequence of values $\{s_1, s_2, ..., s_k, ...\}$ drawn independently from a uniform distribution within the interval $[0, 1]$. At every time step $k$, an input value is injected into one of the nodes of the bosonic chain by resetting one atom state (let us choose site $1$ for the sake of convenience) 
\begin{equation}
|\psi_{k}\rangle =\sqrt{s_k}\left|0\right\rangle +\sqrt{1-s_k}\left|1\right\rangle. \label{eq:input_state}
\end{equation}
 Once, the input $s_k$ has been encoded, the reservoir evolves under the unitary evolution associated with the Hamiltonian in Eq. (\ref{eq:bose-hubbard hamiltonian}). The sequence of input injection and Hamiltonian evolution describes a completely positive trace preserving (CPTP) map of the form

\begin{equation}
\rho_k=e^{-iH\Delta t}\left[\rho_{1,k}\otimes {\rm Tr}_{1}\left\{ \rho_{k-1}\right\} \right]e^{iH\Delta t},
\label{eq:reservoir_evolution}
\end{equation}
where ${\rm Tr}_{1}\left\{ \cdot\right\}$ denotes the partial trace performed over the first site and $\rho_{1,k}=|\psi_{k}\rangle \langle \psi_{k}|$. It is important to notice that time step $\Delta t$ controls the amount of time the information from value $s_k$ propagates through the system. Finally, the output layer consists of measuring a set of $M$ observables $(O_i)$ from the quantum reservoir to extract $M$ features, $\{x_{i}^{(k)}\}_{i=1}^M$, defined as
\begin{equation}
x_{i}^{(k)}={\rm Tr}\left[O_{i}\rho_k\right].
\label{eq:measurment}
\end{equation}

This erase-and-write scheme is based on the original proposal of Fujii and Nakajima (for spins) in Ref. \cite{fujii2017harnessing} and produces a nonlinear input-output mapping where the encoding plays a key role, as discussed in more detail in Ref. \cite{mujal2021analytical,nokkala2021gaussian}. 
The reservoir retains information about past inputs, allowing it to maintain a memory of different input patterns, which is a crucial aspect of QRC that distinguishes it from other approaches like Quantum Extreme Learning Machines. This memory capability is essential for online processing of time-dependent data. 

A comprehensive training dataset, denoted as X, can be constructed by combining all the features, $x_i^{(k)}$, with dimensions $L \times (M+1)$. Here, $L$ represents the number of training inputs  ($s_k$), while $M+1$ accounts for the number of observables plus a constant bias term. In the tasks analyzed in the following we will consider an output layer consisting of two sets
of observables $\langle a_{i}^{\dagger}a_{j} +h.c.\rangle$ and $\langle a_i^\dagger a_i a_j^\dagger a_j\rangle$ for all sites. Additional features can be extracted from the dynamics by leveraging the temporal multiplexing technique \cite{fujii2017harnessing}. Specifically, instead of measuring an observable at time $\Delta t$, the unitary evolution can be divided into $V$ time intervals. This enables a larger dataset with dimensions $ L \times (VM+1)$, where $V$ is the number of virtual nodes  \cite{martinez2023information}. Unless otherwise stated, in all results presented, a total of $V=10$ virtual nodes are employed.

The learning stage consists of three phases: wash-out, training, and testing, of different lengths. In the wash-out step, we evolve the system during a transient of $100$ time steps, in order to mitigate the influence of initial conditions on the system's performance, thereby ensuring the fulfillment of the echo state property \cite{jaeger2001echo, llodra2023benchmarking}. Subsequently, a linear regression model 
\begin{equation}
y_k = \sum_i w_i x_i^{(k)} + b_k, 
\end{equation} 
is trained during $L=1000$ time steps.
Here $w_i$ and $b_k$ are optimized to minimize a Ridge regression loss function with a regularization parameter of $\beta=10^{-2}$ \cite{hoerl1970ridge,beck1977parameter}. It is crucial to note that parameters in the reservoir are fixed and only the output layer parameters are optimized during this step. Finally, the system is tested in $1000$ time steps, making predictions in different tasks $\widehat{y}_k$ and evaluating the model's performance (i.e. the ability of $ {y}_k$ to approximate $\widehat{y}_k$ for unseen inputs) with the optimized parameters. In Sec. \ref{sec:Reservoir performance} we will present results for different tasks, considering expectation values on an infinite number of reservoirs, while in Sec. \ref{sec:measurement_effects} we analyze the performance beyond ideal conditions, with a finite ensemble of identical reservoirs, leading to statistical errors in the measured quantities at the output layer \cite{mujal2023time,hu2023tackling}.

\subsection{Reservoir in different regimes}\label{sec:analytical-Mott}

Before addressing specific tasks, let us consider the limiting regimes of the Bose-Hubbard reservoir. With respect to previous works of bosonic QRC in harmonic networks \cite{nokkala2021gaussian,nokkala2024retrieving,garcia2023scalable,dudas2023quantum},  a Bose-Hubbard reservoir displays interactions, that can expand the capabilities of QRC beyond Gaussian states for machine learning purposes (as started to be explored in QRC \cite{kalfus2022hilbert} and other variational settings \cite{stornati2023variational}). However, the interplay with many-body effects also plays a key role. We advance that we find (see section \ref{sec:Reservoir performance}) that in the presence of very strong interactions, the QRC performance is actually hindered
in the present setting. In the limit case $J \ll U$, when the interaction term completely dominates, Eq. (\ref{eq:bose-hubbard hamiltonian}) is approximated by $H_I = \frac{U}{2}\sum_{j=1}^{N}n_j(n_j-1)$, and the ground state exhibits the Mott-Insulator phase. Then, the dynamics of each node in the network becomes decoupled from the others and the locally encoded information does not propagate through the system. 
Consequently, the information injected at iteration $k$ into the first node in Eq. (\ref{eq:reservoir_evolution}) is completely erased at the next time step, $k+1$. Therefore, the system has no memory and is unable to perform temporal tasks. This regime extends also for finite, yet small values of $J/U$, where the system still exhibits localization, which has been reported to be a hindering factor for QRC \cite{martinez2021dynamical,xia2022reservoir}.

In the limit case $J \gg U$, the system becomes a Bose-Einstein condensate, described by the coupling term $H_C=-J\sum_{j=1}^{N}(b^\dagger_j b_{j+1} + \mathrm{h.c.})$. In this superfluid regime, information encoded into the first node rapidly propagates to the other nodes. This can be seen for the homogeneous open chain by diagonalizing $H_C$ using the Fourier transformation, which yields a ground state $|\psi \rangle_{C} \propto (\sum_{j=1}^N b_{j}^{\dagger})^{N_e}|0\rangle$, where $N_e$ are the number of excitations over the entire lattice \cite{RevModPhys.80.885,Greiner2002}. In this regime, the eigenstates are delocalized in real space, facilitating the propagation of information across the network. The delocalization feature ensures that local erasure of information does not hinder information storage (that can be present globally), so the system maintains memory of the inputs over time, making the QRC algorithm effective. Furthermore, the reservoir operation for a complex bosonic network, in the case of $U = 0$ and a fully connected random network has already been reported to operate as a QRC \cite{llodra2023benchmarking}.

Thermalized quantum reservoirs near the phase transition boundary have been found to excel in nonlinear learning tasks, for disordered spin systems \cite{martinez2021dynamical,%}. Xia \textit{et al.} \cite{
xia2022reservoir}. 
In contrast, our study reveals that a one-dimensional bosonic reservoir can also enhance learning capabilities near the phase transition, without requiring disorder. Furthermore, we establish a direct link between performance and the underlying physics of the Hamiltonian demonstrated by the $\langle r \rangle$ metric (Fig. \ref{fig:ratio_and_gfd}a) and more notably by the $\langle \tilde{D}_1 \rangle$ metric (Fig. \ref{fig:ratio_and_gfd}b) \cite{ kolovsky2004quantum,pausch2021chaos, pausch2021chaos_NJP}, still unexplored in the context of QRC.
Most work on QRC considers disordered reservoirs, which are expected to increase the expressivity of the system. In contrast, regular lattices exhibit symmetries that may reduce the linear independence of the output layer features. However, our conjecture is that the dynamical phase sets the capability of the system to process information being disorder less relevant, at least in small reservoirs. 

In the following, we will address different QRC tasks considering the Bose-Hubbard reservoir for different relative strengths of interaction and tunneling and selecting three representative regimes for the Mott-Insulator ($J/U=10^{-3}$), chaotic ($J/U=0.1$), and superfluid ($J/U=10^{3}$) regimes. We will then compare different topologies, such as periodic, open, and disordered atomic chains.

\section{Reservoir performance}
\label{sec:Reservoir performance}
In this section, we assess the capabilities of a one-dimensional reservoir, either homogeneous or not, with open boundary conditions through a series of benchmarking tasks, including Short-Term Memory (STM), Parity Check (PC), and the Nonlinear AutoRegressive Moving Average (NARMA). These tasks serve as common indicators for evaluating two ingredient components to process temporal information: memory and nonlinearity. 

\subsection{Short-Term Memory task}\label{sec: STM}

The first task consists of evaluating the memory capacity through the STM task, which displays  %. Specifically, we assess 
the system's ability to retain information from previously injected inputs. Specifically, 
an input $s_k$ is injected into the system at each step $k$ and the target is the previous one for a given delays ($\tau$) %with the following target function
\begin{equation}
    \widehat{y}_k = s_{k-\tau}.
    \label{eq:STM-target}
\end{equation}
Ultimately, the memory capacity is evaluated through a correlation, specifically the square of the Pearson factor, between the output, represented by $y_k$, and the target $\widehat{y}_k$:
\begin{equation}
    C\text{=}\frac{{\rm Cov}^{2}\left(\widehat{y}_{k},y_{k}\right)}{\sigma^{2}(y_{k})\sigma^{2}(\widehat{y}_{k})},
    \label{eq:Memory-capacity}
\end{equation}
where $\sigma(\cdot)$ is the standard deviation and ${\rm Cov(\cdot)}$ indicates the covariance. The metric $C$ is constrained between $C=0$ (no ability to recall past inputs) and $C=1$ (perfect recall of past inputs).

\begin{figure}[t] % Place the figure at the top of the page
    \centering
    \includegraphics[width=0.48\textwidth]{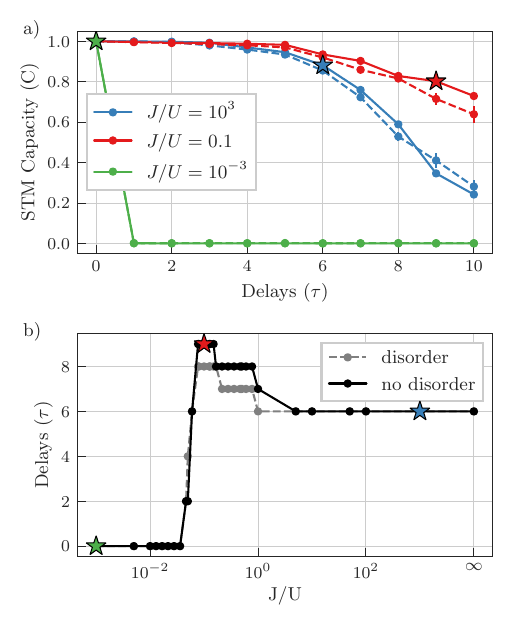} % Adjust the width as needed
    \caption{Performance analysis of the short-term memory task changing the coupling strength. (a) Memory capacity as a function of the input delay for dynamical regimes: Mott-Insulator (green), quantum chaos (red), superfluidity (blue). (b) Maximum delay achieved with a memory capacity above 0.8. The solid (dashed) line represents a homogeneous (heterogeneous with disorder $\delta=0.3$) open-chain of $N=5$ sites with a cut-off $n_c=3$. The reservoir system is trained and tested with 1000 time steps after a wash-out time of 100 (500) time steps if $J/U \geq0.1$ ($J/U < 0.1$). The evolution time ($\Delta t \in [1, 10]$) has been optimized to achieve the best performance. The output layer consists of two sets of observables $\langle a_{i}^{\dagger}a_{j} +h.c.\rangle$ and $\langle a_i^\dagger a_i a_j^\dagger a_j\rangle$, number of virtual nodes $V=10$, ridge regression $\beta=0.01$, initial state of the reservoir all zeros states. For the disordered system, we have averaged over 10 realizations and show the standard error of the mean as the statistical error, in most of the cases below the sign of the symbol.}
    \label{fig:MC_degree_1}
\end{figure}

Let us start our analysis by looking at the solid lines in Fig. \ref{fig:MC_degree_1}a. Here, we analyze how memory capacity decays as we increase the delay across different dynamical phases: Mott-Insulator (green), quantum chaos (red), and superfluidity (blue). To simulate the dynamics numerically, we apply a cut-off of $n_c=3$ on each site, which is sufficient to accurately describe the exact evolution of the system,  as shown in Appendix \ref{app:cut-off}. As expected from the discussion in Sec. \ref{sec:analytical-Mott}, the system lacks memory in the Mott-Insulator phase because injected information does not spread and is wiped out in the next time step. For $\tau = 0$ the memory capacity is maximum ($C=1$), but as soon as a new signal $s_{k+1}$ is introduced, the memory capacity drops to $C=0$ for $\tau \geq 1$ (green line in Fig. \ref{fig:MC_degree_1}a). As $J/U$ increases, the system becomes better at recalling previous inputs for longer times, achieving a memory capacity of $C\geq 0.8$ at $\tau=9$ ($\tau=6)$ in the quantum chaos (superfluidity) regime.

In particular,  for $J/U=0.1$ (red line in Fig. \ref{fig:MC_degree_1}a) the system shows better memory capacity than for $J/U=10^3$ (blue line). To clarify whether the quantum chaotic regime has an advantage over the superfluidity regime in terms of memory capacity, Fig. \ref{fig:MC_degree_1}b plots the maximum delay ($\tau$) with an STM capacity ($C$) greater than $0.8$ for different values of $J$ (star-marked points). Fig. \ref{fig:MC_degree_1}b shows a peak in memory capability within the quantum chaotic regime relative to the superfluidity regime. For each value of $J$ we have tuned the time step $\Delta t \in [1, 10]$ to achieve the best possible results. 

In Fig. \ref{fig:MC_degree_1}b we also compare the case of a homogeneous chain (solid line) with the case of a heterogeneous one (dashed line), which shows comparable performance. For the heterogeneous chain, we have introduced disorder in the coupling term (see Eq. (\ref{eq:bose-hubbard hamiltonian})). Namely, each coupling term is defined as $J_{j, j+1} \in [J_{min}, J_{max}]$ drawn from a random uniform distribution where $J_{min} = J(1 - \delta)$ and $J_{max} = J(1 + \delta)$. We have considered a disorder level of $\delta=0.3$ and averaged the results over 10 realizations to ensure statistical significance. Looking at both, the STM capacity at different delays and the performance across regimes,  disorder does not play a significant role, with heterogeneous chains displaying even less capacity than homogeneous ones.

\begin{figure}
    \centering
    \includegraphics[width=0.48\textwidth]{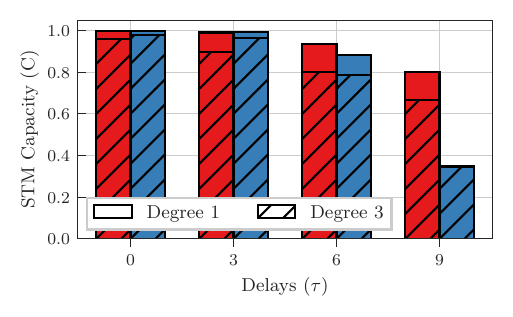}
    \caption{Linear and nonlinear memory capacity as a function of the delay with degrees $d=1$ and $d=3$. The optimal evolution time for the quantum chaos regime (red) is $\Delta t=10$, while for the superfluidity regime (blue) it is $\Delta t = 1$. All other parameters remain consistent with those in Fig. \ref{fig:MC_degree_1}. At $\tau=9$ both performances reach $C=0.35$ after rounding up the third decimal point.}
    \label{fig:MC_degree_123}.
\end{figure}

Let us now study the ability of our system to remember a nonlinear function of an input $s_k$, so the target function is defined as
\begin{equation}
    \widehat{y}_k = s_{k-\tau}^d
    \label{eq:STM-target-nonlinear}
\end{equation}
where $d$ represents the degree of the nonlinearity of the input. Fig. \ref{fig:MC_degree_123} presents a comparative analysis between the linear and the nonlinear case (cubic, $d=3$) by elucidating how distinct dynamical regimes recall nonlinear input functions. In general, we find that the memory capacity decreases with the nonlinearity. Notably, disparities emerge in the performance decay between the superfluidity and quantum chaos regimes. In the superfluidity regime (blue), the memory capacity remains close to one for low delays before exhibiting a pronounced decay with higher delays. Conversely, in the chaotic regime (red), the memory capacity slightly drops at first, especially if we compare the difference between $d=1$ and $d=3$ but for large delays, it clearly outperforms the superfluidity regime for a linear ($d=1$) and a nonlinear ($d=3$) task. Once again, in the Mott-Insulator regime, there is no memory capacity for $\tau \geq 1$.

\subsection{Parity Check task}

\begin{figure}
    \centering
    \includegraphics[width=0.48\textwidth]{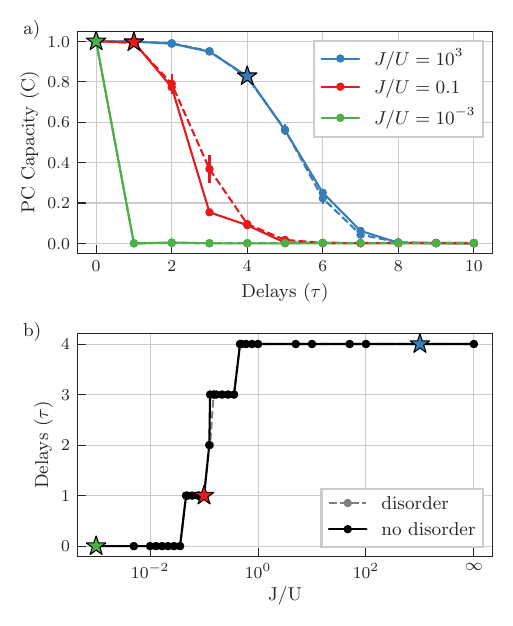}
    \caption{Performance analysis of the parity check task. (a) Memory capacity as a function of the delay for a homogeneous (solid) and inhomogeneous (dotted) one-dimensional chain with open-boundary conditions. (b) Performance for different values of $J/U$ during the transition Mott-Superfluidity, signaling the maximum delay achieved with memory capacity above 0.8. All remaining parameters are consistent with those in Fig. \ref{fig:MC_degree_1}.}
    \label{fig:PC}
\end{figure}

The model's ability to process highly nonlinear information is displayed considering the Parity-Check (PC) task. For this task, we inject a binary random input ($s_k \in \{0, 1\}$) and train the system to fit the following target function
\begin{equation}
    \widehat{y}_k = \sum_j ^\tau s_{k-j} \pmod{2}.
    \label{eq:PC-target}
\end{equation}
Looking at Fig. \ref{fig:PC}, we can see that the outcome of the parity check task differs notably from the short-term memory task (Fig. \ref{fig:MC_degree_1}). In Fig. \ref{fig:PC}a, the performance of the quantum chaos regime drops quickly, but the superfluidity regime maintains a PC capacity above $0.8$ for longer delays. Exploring all regimes there is no peak in Fig. \ref{fig:PC} unlike in the STM task (Fig. \ref{fig:MC_degree_1}), where performance has a maximum near the transition. The lack of such a peak for the parity check is also observed in Xia \textit{et al.} \cite{xia2022reservoir} (check their Fig. 3) and in Martínez-Peña \textit{et al.} \cite{martinez2021dynamical} (Fig. 9)
for the nonlinear memory indicator known as information processing capacity \cite{dambre2012information}.
This is also foreseen from our Fig. \ref{fig:MC_degree_123}, as increasing the degree of nonlinearity reveals that memory capacity is superior in the superfluidity regime compared to the transition phase, at least for low delays. Interesting, the behavior of the memory capacity for this task, in Fig. \ref{fig:PC}b, looks very similar to the information dimension shown in Fig. \ref{fig:ratio_and_gfd}. This seems to indicate that, besides the chaotic regime, efficient QRC is highly related to the eigenvector spreading over the lattice, also in the presence of conserved quantities. Finally, in this task, we can also observe that the performance of a homogeneous and heterogeneous chain is similar.

\subsection{NARMA task}

The NARMA task is well-suited to assess memory and nonlinearity as it simulates a nonlinear dynamical system with a strong dependence on previous inputs ($s_k$). Initially introduced by Atiya \textit{et al.} \cite{atiya2000new}, the NARMA(n) task, 
for $n>2$, is defined as
\begin{equation}
    \widehat{y}_k = 0.3y_{k-1}+0.05y_{k-1}\sum_{j=1}^n y_{k-j} + 1.5s_{k-n}s_{k-1} + 0.1,
    \label{eq:NARMA_N}
\end{equation}
where $s_k$ is the input, $\widehat{y}_k$ is the target, and $n$ is the maximum delay. On the other hand, the NARMA(2) task reads
\begin{equation}
    \widehat{y}_k = 0.4y_{k-1}+0.4y_{k-1}y_{k-2}+0.6s_{k-1}^3+0.1,
    \label{eq:NARMA_2}
\end{equation}
 
For this task, the input $s_k$ is generated from a random uniform distribution ranging from $0$ to $0.2$ as in Ref. \cite{fujii2017harnessing}. To solve the task, the reservoir computer needs to replicate a quadratic nonlinear function of the input sequence up to the maximum delay $n$, see Eq. (\ref{eq:NARMA_N}) for $n>2$ and Eq. (\ref{eq:NARMA_2}) for $n=2$. 

\begin{figure}
    \centering
    \includegraphics[width=0.48\textwidth]{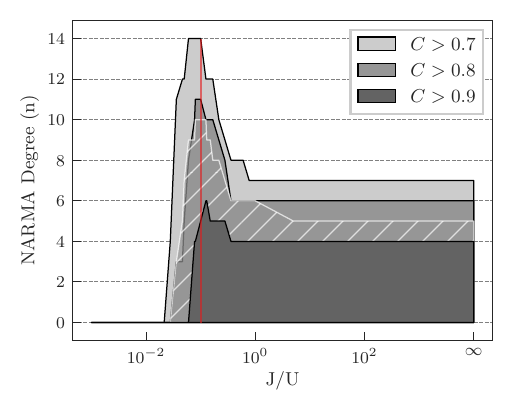}
    \caption{Performance analysis of the NARMA task.  Memory capacity (greyscale) as a function of NARMA (y-axis) and the coupling strength (x-axis). In the quantum chaos phase, there is a notable peak for NARMA(14) with $C \geq 0.7$ and NARMA(5) with $C \geq 0.9$ at $J=0.1$, highlighted by the red line. A. White lines overlapping the $C \geq 0.8$ region represent the averaged results of a heterogeneous chain with disorder ($D=0.3$) over 10 realizations. All remaining parameters are consistent with those in Fig. \ref{fig:MC_degree_1}.}
    \label{fig:NARMA}
\end{figure}

In Fig. \ref{fig:NARMA}, we plot the performance achieved for different NARMA tasks ranging from NARMA(2) to NARMA(14) and for different values of $J/U$. To quantify the capability of the system we set three threshold capacities $C \geq 0.7$ (light grey), $C \geq 0.8$ (grey), and $C \geq 0.9$ (dark grey) to examine how the memory capacity decays as the order of the nonlinearity ($n$) increases. In the Mott-Insulator phase, the capacity never surpasses $C \geq 0.7$. Then, as the system transitions into the quantum chaos phase, a sudden peak is observed, with NARMA(14) exhibiting a memory capacity above 0.7. This peak is gradually reduced, with the system only reaching the $C \geq$ 0.8 and $C \geq$ 0.9 thresholds for NARMA(11) and NARMA(6). Notably, as the system enters the superfluidity regime, the capacity for the three thresholds reaches a plateau, indicating a decline compared to the transition regime in the system's ability to retain memory as the task complexity increases.

In Fig. \ref{fig:NARMA}, we also analyze the effects of disorder. To avoid redundancy, we only present the results for the disordered chain in the case of a performance threshold $C\geq 0.8$. As with the previously discussed tasks, there is no significant difference between using disorder (white) and no disorder (grey), as the memory profile follows a similar pattern across all regimes.

\begin{figure}
    \centering
    \includegraphics[width=0.48\textwidth]{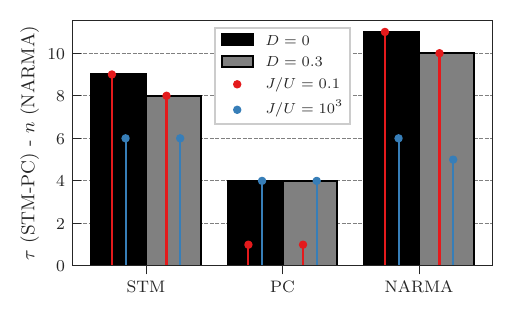}
    \caption{Performance overview of all tasks. The bar height indicates the delay (for STM and PC tasks) and $n$-order  (for the NARMA), with a memory capacity above 0.8. The best performance (maximum delay or order) is displayed, considering all regimes. The blue line represents the maximum capacity in the superfluidity phase while the red one is for the chaotic phase. Performance for both homogeneous (black bars) and disordered (grey bars) chains are displayed.}
    \label{fig:global-1d-open}
\end{figure}

To summarize our main %the principal concepts and illustrate the 
results of different tasks, let us focus on Fig. \ref{fig:global-1d-open}. First, the performance can reach its peak in the quantum chaotic or the superfluid regime, depending on the specific task.
While the quantum chaotic regime often provides the highest performance, as indicated by the bar height consistently matching or exceeding the blue line (representing the superfluidity phase), there are cases where the superfluid regime outperforms the chaotic regime.
Secondly, for a one-dimensional open topology, the introduction of disorder in the coupling strength does not significantly enhance performance.  
 Notice that the black bar (no disorder) is always equal to or higher than the grey bar (disorder). Our hypothesis is that disorder can confer advantages to systems with high symmetry, such as one-dimensional periodic systems.  However, even with disorder, a one-dimensional periodic system can not surpass the capacity of a homogeneous one-dimensional open system.  For further details, we direct the interested reader to section \ref{sec:topologies}.

\section{Measurements effects}
\label{sec:measurement_effects}

This section discusses the limits imposed by statistical noise on the previous benchmarking tasks. Concretely, we compare the performance of the STM under a finite number of measurements ($N_m$) and the ideal case ($N_m \to \infty$) in Sec. \ref{sec:Reservoir performance}.
In real experiments, the ideal expected value of any observables ($x_i^{(k)}$) is affected by the stochastic nature of quantum mechanics, which causes the appearance of statistical errors ($\xi_{N_m}$) that result in an experimental expected value ($x_{i, exp}^{(k)}$) deviating from the ideal $x_i^{(k)}$ leading to:
\begin{equation}
    x_{i, exp}^{(k)} = x_i^{(k)} + \xi_{N_m}.
    \label{eq:noisy_feature}
\end{equation}

To simulate such a stochastic effect, numerically, we add random noise ($\xi_{N_m}$) drawn from a Gaussian distribution with mean $\mu=0$ and standard deviation $\sigma = \frac{1}{\sqrt{N_m}}$. According to the central limit theorem, we expect that for $N_m \gg 1 $, the experimentally measured value will converge to the ideal expected one (corresponding to an infinite ensemble) \cite{palacios2024role,mujal2023time,garcia2023scalable}.

\begin{figure}
    \centering
    \includegraphics[width=0.48\textwidth]{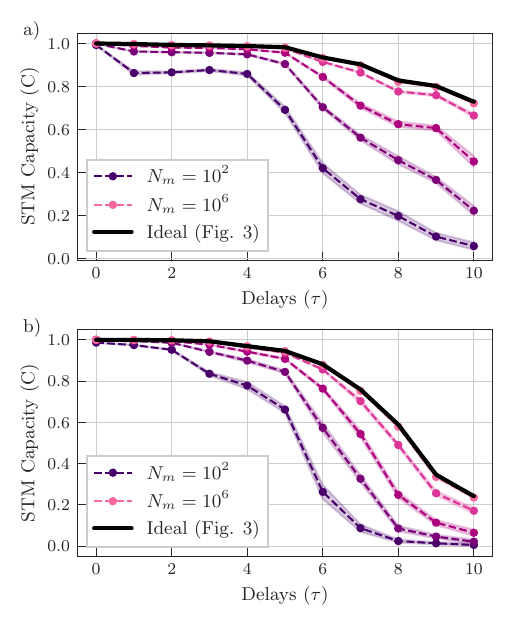}
    \caption{Performance analysis of the STM task as a function of the delay and the number of measurements ($N_m$). The memory profiles show how the model's capacity improves as $N_m$ increases for the (a) quantum chaos and the (b) superfluidity case, $J=0.1$ and $J=10^3$, respectively. The results are averaged over 10 realizations and the statistical error is denoted by the standard deviation.}
    \label{fig:measurement_effects}
\end{figure}

As shown in Fig. \ref{fig:measurement_effects}, the model's capacity improves significantly with the number of measurements ($N_m$). Notably, the ideal performance depicted in Fig. \ref{fig:MC_degree_1} is achieved for both $J=0.1$ and $J=10^3$ when $N_m=10^6$, as illustrated in Fig. \ref{fig:measurement_effects}a and Fig. \ref{fig:measurement_effects}b, respectively. To ensure consistency in our results, we have optimized the time evolution $\Delta t$ within the range $[1, 10]$ and used the same parameters as in section \ref{sec:Reservoir performance}.

We want to point out that the relationship between the number of measurements and performance is model-specific and can be precisely assessed only in special cases \cite{garcia2023scalable}. While the precision of observable expectation values generally improves with approximately $\frac{1}{\sqrt{N_m}}$ (see Fig. \ref{fig:measurement_effects}), translating this into performance across different tasks is complex.

\section{Topology effects}
\label{sec:topologies}

\begin{figure}
    \centering
    \includegraphics[width=0.48\textwidth]{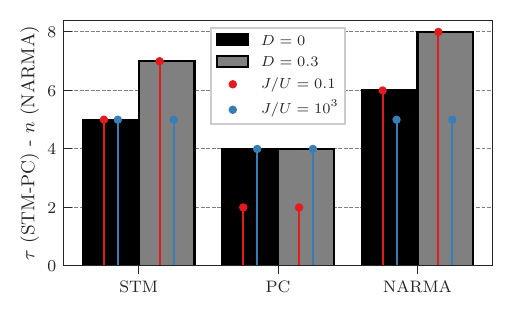}
    \caption{Performance overview of all tasks for a one-dimensional chain with periodic boundary conditions. }
    \label{fig:global-1d-periodic}
\end{figure}

Up to this point, we have focused on a one-dimensional chain of bosons with either homogeneous or heterogeneous coupling strengths. In this section, we extend the analysis to other topologies, beginning with a one-dimensional chain with periodic boundary conditions (PBC). 
In a similar spirit as in Fig. \ref{fig:global-1d-open},  in Fig. \ref{fig:global-1d-periodic} we overview the three performance tasks (STM, PC, and NARMA) for a one-dimensional chain with periodic topology. Notably, Fig. \ref{fig:global-1d-periodic} reveals two key findings. Firstly, comparing Fig. \ref{fig:global-1d-open} and Fig. \ref{fig:global-1d-periodic}, we observe that the periodic chain generally underperforms compared to the open chain. Secondly, in contrast to the previous sections, introducing disorder can improve the model's performance.  Indeed, breaking symmetries such as translational symmetry enables the model to extract novel features, enhancing its learning capabilities.  

\begin{figure}
    \centering
    \includegraphics[width=0.48\textwidth]{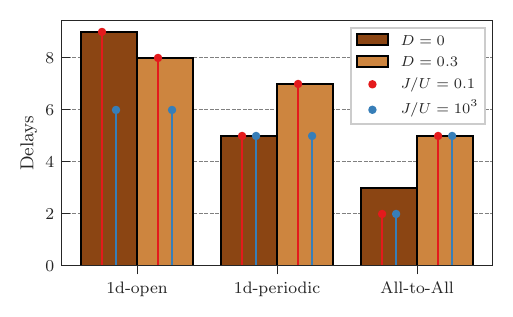}
    \caption{Performance overview of the linear STM task for different topologies of the Bose-Hubbard model.}
    \label{fig:global-STM}
\end{figure}

We now compare the performance of the various topologies in Fig. \ref{fig:global-STM}, also examining the all-to-all interactions, corresponding to a fully complex network. The results indicate that the one-dimensional open topology yields the best results, while the all-to-all topology performs the poorest. A deeper analysis considering the independent degrees of freedom at the output layer indicates that all-to-all network may not be well-suited for quantum reservoir computing, as the features extracted from its quantum dynamics tend to exhibit more redundancy compared to the one-dimensional open topology (see Appendix \ref{app:singular_values} for further details). Across all three topologies, the performance in the quantum chaos phase is either equal to or higher than that observed in the superfluidity regime, as discussed in the previous sections. Although we have chosen the STM with degree one for this analysis, the conclusions would hold similarly if the NARMA task were employed.

\begin{figure}
    \centering
    \includegraphics[width=0.48\textwidth]{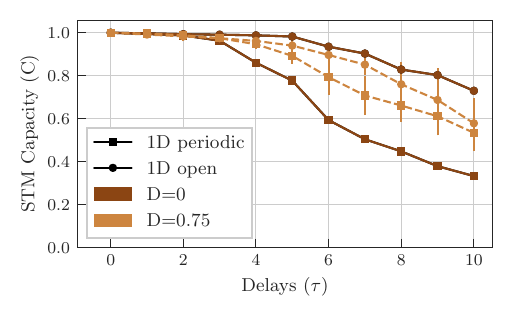}
    \caption{Performance comparison of the linear memory capacity for different topologies. The results are averaged over 10 realizations and the statistical error is denoted by the standard deviation}
    \label{fig:disorder}
\end{figure}

Intuitively, one would expect no performance difference between a periodic and open topology with disorder. However, Fig. \ref{fig:global-STM} reveals that the periodic chain exhibits slightly inferior performance compared to the open chain. In Fig. \ref{fig:disorder}, we observe a significant performance gap between homogeneous topologies. As the coupling disorder increases, the error bars widen, ultimately leading to similar performance for both topologies, as evident from the overlapping error bars. This result suggests that a disorder level of 0.3 is insufficient to completely erase the underlying periodic structure, allowing it to retain a weak form of translational symmetry. This residual symmetry implies that the periodic chain still preserves some level of uniformity across sites, even under moderate disorder. In fact, as disorder strength increases, the results will converge towards those of the open topology. Conversely, with very weak disorder, the periodic topology will retain its performance characteristics, similar to the case with no disorder, which explains the inferior performance in Fig. \ref{fig:global-STM}.   

\section{Conclusions}
\label{sec.conclusions}

We investigated QRC performance across three distinct dynamical regimes. First, and as expected, the Mott-Insulator phase suitability for information processing is limited, as evidenced by STM, PC, and NARMA task results. The quantum chaotic phase where the interplay of coupling and interaction is balanced is instead suited for QRC consistently with previous works \cite{martinez2021dynamical,xia2022reservoir} and demonstrated superior performance in tasks demanding a balance of linear and nonlinear memory. Finally, the superfluid regime, when interactions are negligeable, also yielded excellent outcomes for nonlinear tasks. 
This comes as a bit of a surprise, as this phase is not fully ergodic and this could be expected to limit the expressivity. 
By extending the results of \cite{martinez2021dynamical}, we found here that the generalized multifractal dimension is a fundamental tool to relate the QRC performance to the dynamical phases of the reservoir.
Our results suggest that efficient information propagation within the system, rather than proximity to a phase boundary, is the critical factor for QRC's success based on local input injection.

Furthermore, our topology analysis of one-dimensional chains with open and periodic boundary conditions, as well as all-to-all connectivity, revealed a surprising finding: disorder is unnecessary in systems lacking translational symmetry. The one-dimensional chain with open boundaries consistently outperformed other topologies across various benchmarks. These results challenge the conventional wisdom and suggest that more complex topologies should be explored, and prioritize OBC over PBC or all-to-all connectivity in future QRC studies. It will be interesting to test the validity of these conclusions in larger reservoirs.

To conclude, our results establish Bose-Habbard chains as suitable QRC, not only in the chaotic phase but also in the superfluid one, and offer new insights into the design of more straightforward and efficient QRC implementations using homogeneous many-body reservoirs \cite{PhysRevB.109.064202}. This work represents a promising step towards more practical experimental applications in quantum information processing, particularly for systems operating on near-term quantum devices. The results highlight the importance of selecting appropriate dynamical regimes and topologies in QRC, suggesting a path toward optimized performance without relying on complex disordered systems. 

\begin{acknowledgments}

We acknowledge insightful discussions during the Interantional Workshop on Disordered Systems 2024, held in Salamanca, with Andreas Buchleitner, Rafael Molina, Manuel Pino, and  Alberto Rodríguez. 
We acknowledge the Spanish State Research Agency, through the Mar\'ia de Maeztu project CEX2021-001164-M funded by MCIU/AEI/10.13039/501100011033, through the COQUSY project PID2022-140506NB-C21 and -C22 funded by MCIU/AEI/10.13039/501100011033, and the QuantERA QNet project PCI2024-153410 funded by MCIU/AEI/10.13039/501100011033 and cofounded by the European Union; MINECO through the QUANTUM SPAIN project, and EU through the RTRP - NextGenerationEU within the framework of the Digital Spain 2025 Agenda. The CSIC Interdisciplinary Thematic Platform (PTI+) on Quantum Technologies in Spain (QTEP+) is also acknowledged.
This project has received funding from MICIN and Generalitat de Catalunya with funding from the European Union, NextGenerationEU (PRTR-C17.I1),
the Government of Spain (Severo Ochoa CEX2019-000910-S and FUNQIP), Fundació Cellex, Fundació Mir-Puig, Generalitat de Catalunya (CERCA program).
\end{acknowledgments}

\appendix

\renewcommand{\thefigure}{A\arabic{figure}}
\setcounter{figure}{0}
\section{Singular values}
\label{app:singular_values}

As shown in Section \ref{sec:topologies}, the homogeneous all-to-all topology is the least suitable for QRC due to the redundant features. In contrast, systems with no translational symmetry, such as the one-dimensional open topology, offer more promising opportunities for QRC. In this section, we will quantify the redundancy by analyzing the singular values of the training dataset ($X_{train}$) from Eq.  (\ref{eq:measurment}).

\begin{figure}[h]
    \centering
    \includegraphics[width=0.48\textwidth]{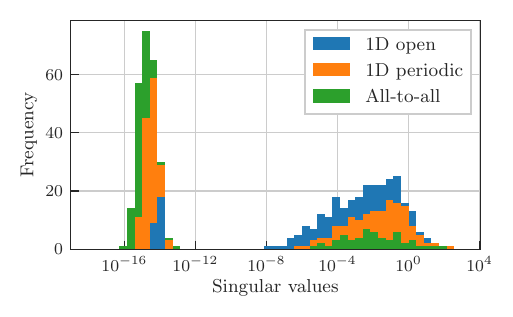}
    \caption{Histogram with all the singular values from the $X_{train}$ used in Fig. \ref{fig:global-STM} for D=0.}
    \label{figB1:svd_topologies}
\end{figure}

In Fig. \ref{figB1:svd_topologies}, we notice a striking difference in the distribution of singular values across various topologies. The all-to-all topology exhibits the largest number of singular values near zero, indicating a high degree of redundancy in the dataset. This redundancy is consistent with the poor performance observed in Fig. \ref{fig:global-STM}. In contrast, the one-dimensional topology exhibits a majority of non-zero singular values, suggesting that the features are less redundant and may provide better results on unseen data. This difference in redundancy highlights the importance of topology selection in achieving optimal performance.

\section{Validty of numerical results}
\label{app:cut-off}

\begin{figure}
    \centering
    \includegraphics[width=0.48\textwidth]{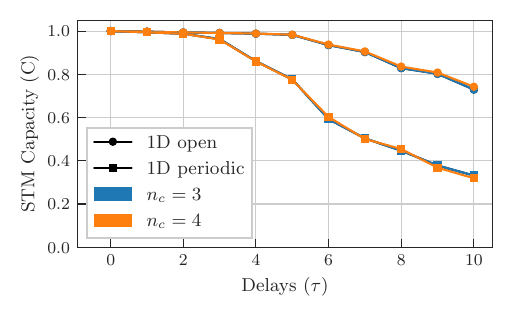}
    \caption{Performance analysis of the STM task for different cut-off $n_c=3$ (blue) and $n_c=4$ (orange) for $J=0.1$.}
    \label{figD1:boson-cutoff}
\end{figure}

The input injection scheme utilized to implement the QRC implies a nonconservation of the number of excitations in the reservoir. Consequently, an exact simulation would require spanning the infinite-dimensional Hilbert space.  However, we have checked that, in the operational regime employed, the number of excitations was always very limited. In fact, all the results presented in the main text have been obtained using a Fock-space cut-off of $n_c=3$ for each system site. To assess the validity of this approximation, in Fig. \ref{figD1:boson-cutoff} we evaluate the performance of the STM capacity for both the open and periodic chains, comparing the truncation level $n_c=3$ with $n_c=4$. The results demonstrate that the truncation has minimal impact on the outcome, indicating that the reservoir dynamics are accurately reproduced.

\bibliographystyle{unsrt}
\bibliography{main}

\end{document}